\documentstyle[11pt,paspconf,epsf]{article}

\begin{document}

\title{Cluster deprojection with joint lensing, X-ray, and 
       Sunyaev-Zeldovich data}

\author{Katrin Reblinsky and Matthias Bartelmann}
\affil{Max Planck Institut f\"ur Astrophysik, P.O. 1523, D-85740 Garching,
Germany}

\begin{abstract}
We propose a new cluster deprojection algorithm for recovering cluster
structure along the line of sight (los) based on the Richardson-Lucy (RL) 
algorithm, which recovers a nonnegative theoretical distribution 
from a given projection.
To optimize our reconstruction we combine X-ray and 
Sunyaev-Zeldovich (SZ) data with gravitational
lensing maps. The combination of these three data types constrains 
the structure of rich clusters of galaxies.
To validate our new approach, we apply the algorithm to gas--dynamical 
cluster simulations. A reconstruction using the lensing potential is shown.

\end{abstract}


%
%
\section{Richardson-Lucy deprojection algorithm}
Resolving the los-structure of 3d objects can be formulated as 
inversion problem,
\begin{displaymath}
\psi(x) = \int \phi(\zeta) P(x|\zeta) \,\mbox{d}\zeta,
\end{displaymath}
where $\phi(\zeta)$ is the function of interest, $\psi(x)$ is the function
accessible through measurement, and the integral kernel $P(x|\zeta)$ is 
normalized and non--negative as $\psi(x)$ and $\phi(\zeta)$ represent
probability distribution functions.

Using Bayes theorem for conditional probabilities, a two step 
{\em iterative reconstruction algorithm} (Lucy, 1974; Lucy, 1994) 
\begin{displaymath}
\psi^{r}(x)=\int \phi^{r}(\zeta) P(x|\zeta) \,\mbox{d} \zeta
\hspace{2cm}
\phi^{r+1}(\zeta)=\phi^{r}(\zeta) \int 
\frac{\psi(x)}{\psi^{r}(x)} P(x|\zeta) \,\mbox{d}x
\end{displaymath}
for approximating $\phi(\zeta)$ can be derived, 
with $\phi^{r+1}$ being the $(r+1)$-st estimate.
For systems with axial symmetry inclined with respect to the los
an explicit expression for $P(x|\zeta)$
was derived by Binney et al. (1990).

\section{The proposed algorithm}
The three observables depend differently on the los-structure of the 
3d gravitational potential (Reblinsky \& Bartelmann (1999)):
\begin{displaymath}
\mbox{Lensing potential}\quad
\psi(x,y) \propto \int_{-\infty}^{+\infty} \phi(R,Z) \mbox{d}z,
\end{displaymath}
\begin{displaymath}
\mbox{X-ray emission}\quad
S_x(x,y) \propto \int_{-\infty}^{+\infty} 
\sqrt{T} \exp{[-2 \phi^{\prime}(R,Z)]} \mbox{d}z,
\end{displaymath}
\begin{displaymath}
\mbox{SZ decrement}\quad
\Delta T_{\mbox{\footnotesize SZ}}(x,y) \propto \int_{-\infty}^{+\infty} 
T \exp{[- \phi^{\prime}(R,Z)]} 
\mbox{d}z.
\end{displaymath}

For the X-ray case and the SZ-effect an isothermal gas distribution is
assumed. The inclination angle $i$ and the temperature $T$ are input
parameters.
This leads to the following {\em combined two step reconstruction 
formula:}
\begin{eqnarray*}
\mbox{Step 1:} &\quad&
F_{n_i} = \int f_{n_i}(x,y)P(x,y|R,Z)\,\mbox{d}x\mbox{d}y \\
\mbox{Step 2:} &\quad&
\phi_{r+1} \propto \phi_r 
           \left[
             \alpha F_{n_1} 
            +\beta \left(1-\frac{1}{2}\ln F_{n_2}\right)
            +\gamma\bigg(1-\ln F_{n_3}\bigg)
           \right],
\end{eqnarray*}
where the $f_{n_i}$ are defined as
$f_{n_1}=\frac{\psi}{\psi_n}$,
$f_{n_2}=\frac{S_x}{S_{x,n}}$,
and $f_{n_3}=\frac{\Delta T_{\mbox{\footnotesize SZ}}}{\Delta
T_{\mbox{\footnotesize SZ},n}}$.

\section{Application to test data from gas--dynamical cluster simulations}
Using the lensing potential $\psi(x,y)$ as input data, the
characteristic features of the true 3d potential $\phi(R,Z)$ 
are recovered very well after 7-8 iteration cycles. 
We also find that the reconstruction is insensitive to the initial guess,
even $\phi(R,Z)=const$ can be used as in Fig. 1.
\begin{figure}
\plottwo{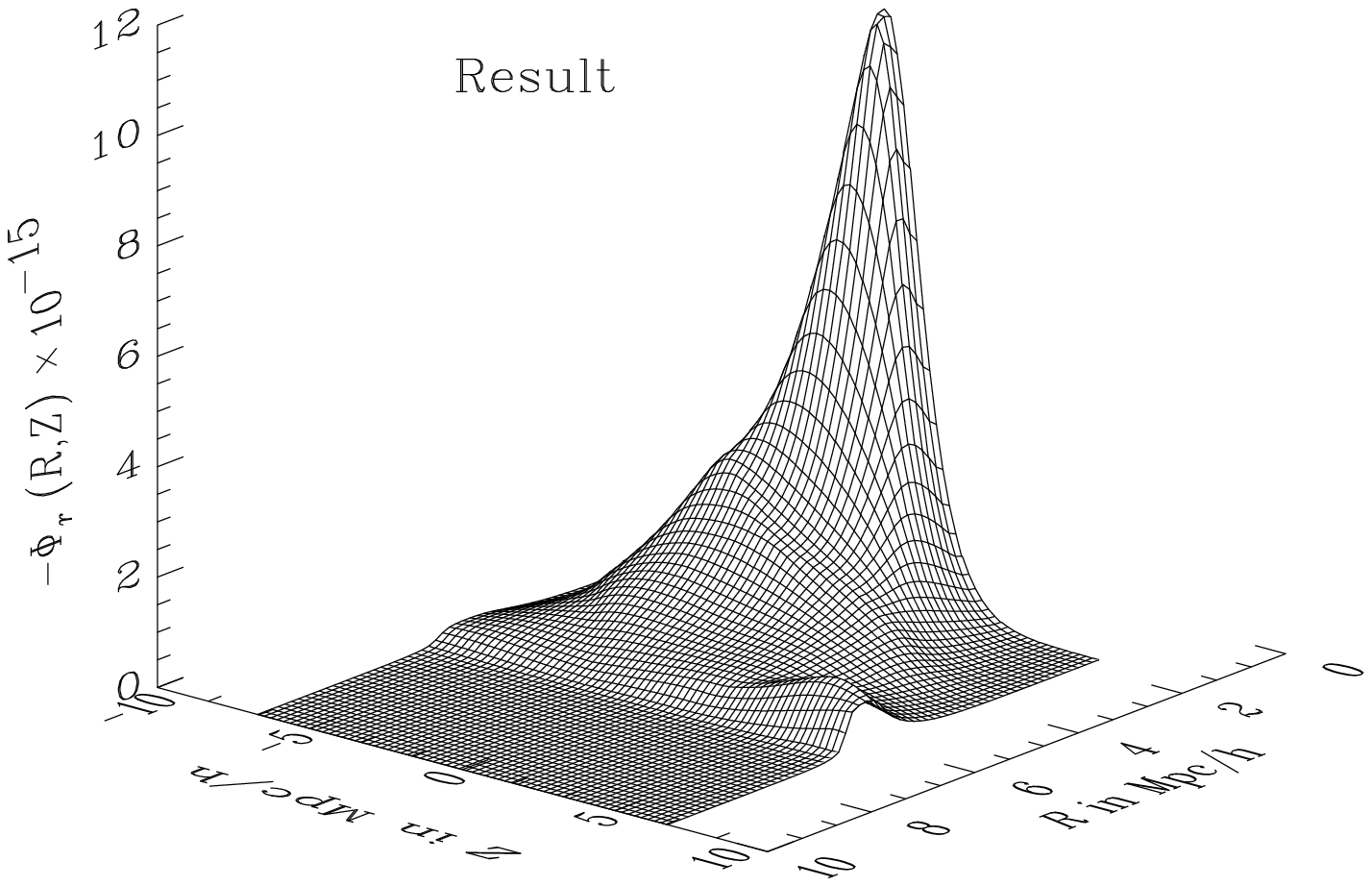}{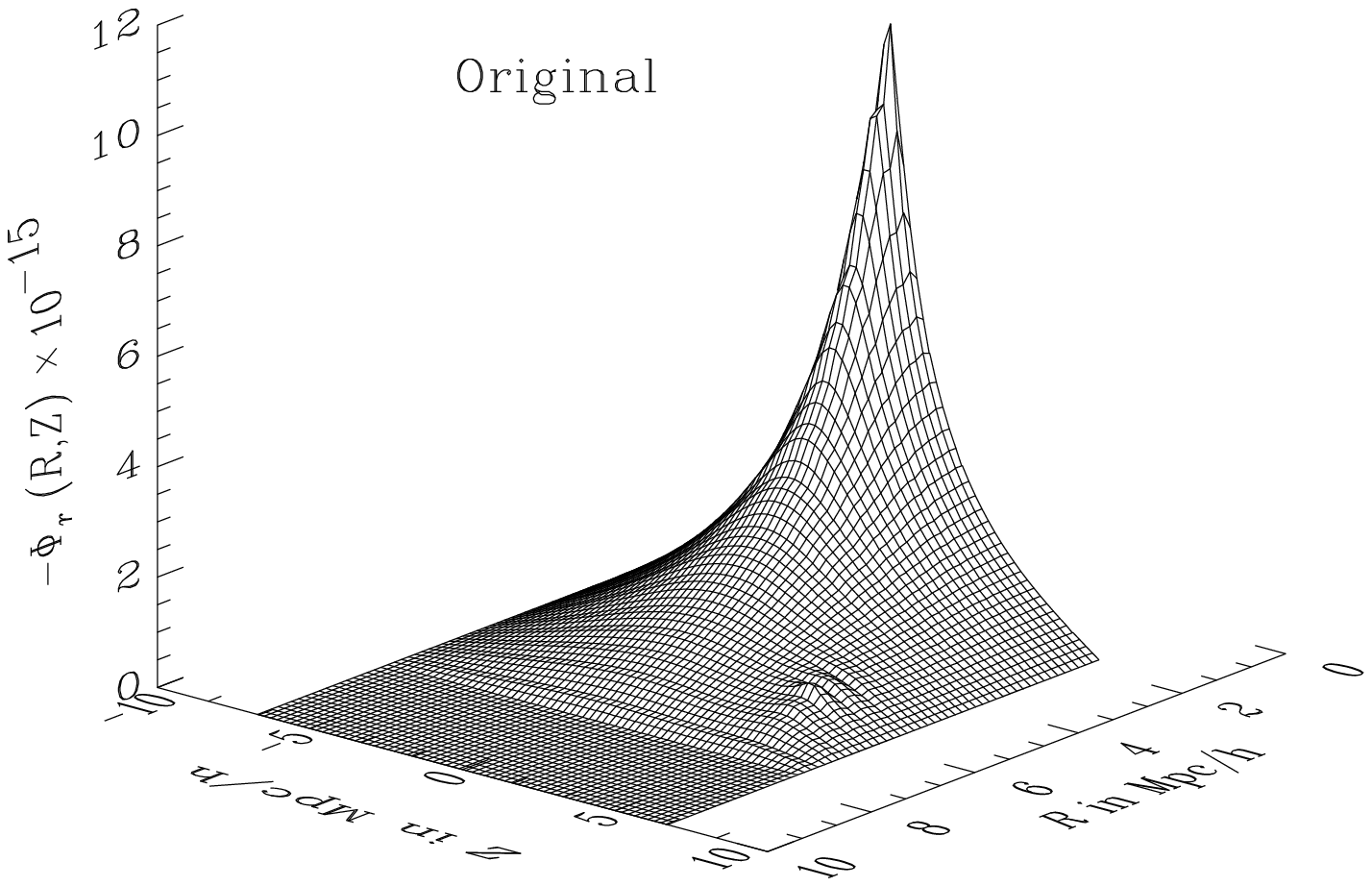}
\caption{Reconstruction of the 3d gravitational potential $\phi$ (displayed is 
$-\phi$) from the lensing potential. Inclination angle $i=30^{\circ}$. 
Left: the result after 7 iteration cycles; Right: the original cluster potential
used to compute the input data, i.e. the lensing potential $\psi$.}
\end{figure}

\acknowledgements

This work was supported in part by the Sonderforschungsbereich 375 of the
Deutsche Forschungsgemeinschaft.

%
%

\end{document}